\documentclass{cpbtex}
\usepackage{graphicx}
\begin{document}
\begin{CJK*}{GBK}{song}

\title{Equilibrium Dynamics of the Sub-Ohmic Spin-boson Model At Finite Temperature \thanks{This work is supported by NSFC grants (11374362, 11974420), Fundamental Research Funds for the Central Universities, and the Research Funds of Renmin University of China 15XNLQ03. }}


\author{Ke Yang$^{1}$ and \ Ning-Hua Tong$^{1}$\thanks{Corresponding author. E-mail: nhtong@ruc.edu.cn}\\
$^{1}${ Department of Physics, Renmin University of China, 100872 Beijing, China}  \\  
}

\date{\today}
\maketitle

\begin{abstract}
We use the full-density matrix (FDM) numerical renormalization group (NRG) method to calculate the equilibrium dynamical correlation function $C(\omega)$ of the spin operator $\sigma_z$ at finite temperature for the sub-Ohmic spin-boson model. A peak is observed at the frequency $\omega_{T}\sim T$ in the curve of $C(\omega)$. The curve merges with the zero temperature $C(\omega)$ in $\omega \gg \omega_{T}$ and deviate significantly from the power-law form $\omega^{\pm s}$ of the zero temperature curve in $\omega \ll\omega_{T}$. 

\end{abstract}



\section{Introduction}
The spin-boson model (SBM) is widely used in modelling a two-level system interacting with the environmental bath.\ucite{Leggett1,Leggett2,Weiss1} Using the SBM, one can easily build the decoherence\ucite{Alex1} or damping\ucite{Anders1} picture of the quantum impurity system due to influence of the bath. So SBM is frequently applied in contexts such as damping in electric circuits, decoherence of quantum oscillations in qubits,\ucite{Costi1,Khveshchenko1,Thorwart1} thermal conductance,\ucite{Ruokola1} {\it etc.}. It has also been used in quantum information processing.\ucite{Chin2} The coupling between the spin and bosonic bath is described by the spectral function $J(\omega) \sim \alpha \omega^s$. In this paper, we focus on the sub-Ohmic case with $0 <s <1$ in the delocalized side ($\alpha < \alpha_c$) of the delocalize-localize quantum phase transition (QPT).  

The spin-boson model has been studied by many theoretical as well as numerical techniques in the past few decades. Although an exact solution has not be obtained, various methods, such as path integral method,\ucite{Leggett2} variational method,\ucite{Lv1} bosonic NRG,\ucite{Bulla1,Bulla2} quantum Monte Carlo (QMC),\ucite{Winter1} exact diagonalization,\ucite{Alvermann1} {\it etc.} have deepened our understanding of this model. Among the many properties of this model, the equilibrium spin correlation function $C(\omega)$ (defined below) is of particular interest because it not only reflects the damping effect of the bath on the spin, but also shows nontrivial critical properties close to the QPT. Previous studies of $C(\omega)$ mainly focus on zero temperature.\ucite{Anders1,Kehrein1,Herbert1,Florens1,Zheng1,Zheng2,Zheng3} For the sub-Ohmic case with zero bias $\epsilon=0$ and $\alpha < \alpha_c$, the zero temperature $C(\omega)$ has a peak at the renormalized tunnelling strength $\omega = \Delta_r$. For $\alpha$ close to $\alpha_c$, a crossover frequency $\omega_{cr}$ separate the low frequency regime into two parts: $C(\omega) \sim \omega^{-s}$ for $\omega_{cr} \ll \omega \ll \Delta_{cr} $ and  $C(\omega) \sim \omega^{s}$ for $\omega \ll \omega_{cr}$.\ucite{Kehrein1,Herbert1} $\omega_{cr} \sim T^{\ast}$ is the crossover energy scale between the delocalized fixed point and the critical fixed point.\ucite{Anders1,Florens1} When a bias is exerted, a competition between the energy scales $\epsilon$ and $T^{\ast}$ are observed and the Shiba relation\ucite{Sassetti1} still holds in $C(\omega)$.\ucite{Zheng3} For the spin-boson model with quadratic spin-boson coupling, the study of equilibrium dynamical correlation functions at zero temperature shows that the power-law $\omega$-dependence occurs at and away from the critical point.\ucite{Zheng1,Zheng2}  Much less is known for the behavior of $C(\omega)$ at finite temperature, however. One would expect that the thermal excitation of the SBM would significantly increase the damping of the two-level system and invalidate the Shiba relation in the low frequency regime.\ucite{Anders1} It is the purpose of this paper to present a quantitative study of $C(\omega)$ at finite $T$ and zero bias.
 
The bosonic NRG is a powerful tool to study quantum impurity models including SBM. There are various ways to generate the equilibrium dynamical quantities such as $C(\omega)$ from NRG data, i.e., from the eigenstates and eigenenergies produced in the NRG iterative diagonalization. The patching scheme produces $C(\omega)$ by empirically combining the spectral functions generated from each energy shell of the SBM\ucite{Bulla3} and does not guarantee exact sum rule. The density-matrix NRG (DM NRG) combines the data of each energy shell using the reduced density matrix of the full system, such that the influence of the low energy state on the high frequency spectral function is well described.\ucite{Hofstetter1} Although DM NRG combines NRG data from all NRG iterations, it works through patching scheme. For finite temperature, one still needs to set the temperature according to the energy scale of a chosen shell.\ucite{Weichselbaum1} A true multi-shell framework was built on the full density matrix and the complete basis sets introduced by Anders and Schiller.\ucite{Anders2,Anders4} The FDM NRG treats the density matrix exactly and guarantees the sum rule rigorously.\ucite{Weichselbaum2} Recently, we developed the full excitation (FE) NRG method for calculating the equilibrium dynamical quantities.\ucite{Yang1} It treats the density matrix exactly and employs the full excitations of NRG, i.e., both intra-shell and inter-shell excitations are taken into account. This method guarantees both the sum rule and the positiveness of the diagonal spectral function. In this paper, we will use FDM NRG to study finite temperature $C(\omega)$ for SBM, since FDM NRG is much faster than FE NRG and with suitable broadening, the problem of negative spectral function of FDM NRG method does not influence our conclusion.

Our results show that a peak emerges at the frequency $\omega_{T} \sim T$ in the curve of $C(\omega)$. For $\omega \gg \omega_{T}$, $C(\omega)$ merges with the zero temperature curve. For $\omega \ll \omega_T$, we observe that $C(\omega)$ deviates significantly from the zero temperature curve, but a reliable quantitative description of $C(\omega)$ in this regime is difficult to obtain, due to the limitation of NRG method. Our work is organized as following: We introduce SBM and NRG method in section II; results of $C(\omega)$ are shown in section III; a final summary is given in section IV.

\section{Model and  Method}

The Hamiltonian of SBM reads
\begin{eqnarray}
H=-\frac{\Delta}{2}\sigma_{x}+\frac{\epsilon}{2}\sigma_{z}+\sum_{i}\omega_{i}a^{\dag}_{i}a_{i}+\frac{\sigma_{z}}{2}\sum_{i}\lambda_{i}(a_{i}+a^{\dag}_{i}).
\end{eqnarray}
Here, $\epsilon$ is the bias and $\Delta$ is the tunnelling strength between the two levels. The spin represented by the Pauli matrices is coupled to a bosonic bath. $\omega_{i}$ is the frequency of the $i$-th boson bath site and $\lambda_{i}$ is the coupling strength between the spin and bath mode $i$. The coupling is encoded into the spectral function
\begin{equation}
   J(\omega) = \pi \sum_{i} \lambda_{i}^{2} \delta(\omega -   \omega_{i}).
\end{equation}
Usually the following power-law form of $J(\omega)$ with a cut-off  $\omega_c=1$ is used,
\begin{equation}
   J(\omega) = 2\pi\alpha \omega^{s} \omega_{c}^{1-s}, \,\,\,\,\, \,\,\,\, (0 \leqslant \omega  \leqslant   \omega_c).
\end{equation}
In this form, the coupling strength is described by $\alpha$. For the sub-Ohmic bath ($0 \leqslant s < 1$) with $\epsilon=0$ and $T=0$, a quantum phase transition occurs at $\alpha = \alpha_c$ between the delocalized state ($\alpha < \alpha_c)$ and the localized state ($\alpha > \alpha_c$).\ucite{Bulla2,Kehrein1}

The NRG calculation consists of three steps, the logarithmic discretization, mapping to Wilson chain, and the iterative diagonalization.\ucite{Wilson1,Bulla4} In the first step, the continuous bath is discretized to bath sites with energies decreasing as $\omega_{n} \propto \Lambda^{-n}$. $\Lambda \geqslant 1$ is the logarithmic discretization parameter which controls the logarithmic discretization error. The continuous bath is recovered only in the limit $\Lambda = 1$.  In the second step, the star-like discretized Hamiltonian is mapped into a Wilson chain with the Hamiltonian\ucite{Bulla2}
\begin{eqnarray}
H_{c}=H_{imp}+\sqrt{\frac{\eta_{0}}{\pi}}\frac{\sigma_{z}}{2}(b_{0}+b^{\dag}_{0})+\sum^{\infty}_{n=0} \left[ \epsilon_{n}b^{\dag}_{n}b_{n}+t_{n}(b^{\dag}_{n}b_{n+1}+b^{\dag}_{n+1}b_{n}) \right].
\end{eqnarray}
The chain Hamiltonian describes an impurity (spin) $H_{imp}$ coupled to the zeroth bath site. The $n$th ($n>0$) boson bath site only interacts with the nearest neighbor bath sites through the hopping $t_{n}$. Both $t_n$ and $\epsilon_n$ decrease as $\Lambda^{-n}$ with increasing $n$. For a finite chain of length $N$, we multiply a factor $\Lambda^{N}$ to rescale the smallest energy scale of the chain into unity. The truncated chain Hamiltonian is then written as
\begin{eqnarray}
H_{N}=\Lambda^{N}[H_{loc}+\sqrt{\frac{\eta_{0}}{\pi}}\frac{\sigma_{z}}{2}(b_{0}+b^{\dag}_{0})+\sum^{N}_{n=0}\epsilon_{n}b^{\dag}_{n}b_{n}+\sum^{N-1}_{n=0}t_{n}(b^{\dag}_{n}b_{n+1}+b^{\dag}_{n+1}b_{n})].
\end{eqnarray}
Diagonalization of $H_{n}$ gives the eigenenergies and eigenstates of the $n$th energy shell. An iterative scheme is used to implement the diagonalization process for a given chain length $N$. We start from diagonalizing the Hamiltonian of a short chain whose Hilbert space dimension is small. Then we add a new bath site to the chain and build the Hamiltonian matrix on the product basis of the obtained eigenstates and the states of newly added site. The matrix is diagonalized again and the next site is added. This process continues until all the sites are added. To avoid the exponential enlargement of the Hilbert space dimension, a truncation of the energy spectrum is introduced after each diagonalization: only the lowest $M_s$ eigenstates are kept and used to form the new Hilbert space. The high energy discarded eigenstates are also stored for computing physical quantities later. This spectrum truncation introduces the NRG truncation error which diminishes in the limit $M_s = \infty$.\ucite{Yang2} For the bosonic NRG, the number of states of each bosonic bath site has also to be truncated. For the newly added site, we use the lowest $N_b$ occupation number states to build the Hamiltonian matrix which has a linear size $M_s \times N_b$. The infinitely large local bosonic Hilbert space is recovered in the limit $N_b=\infty$. We therefore have three NRG parameters $\Lambda$, $M_s$, and $N_b$ to control the numerical error of bosonic NRG. The exact result are obtained only in the simultaneous limit $\Lambda=1$, $M_s=\infty$, and $N_b=\infty$. 
Conclusion from NRG study should be checked by extrapolating the numerical data to the above limit.

Physical quantities can be calculated from the eigenstates $\{\vert{s}\rangle_{n}^{X}\}$ and eigenenergies $E_{s}^{n}$ ($s \in [1, M_s \times N_b]$, $n \in [0, N]$) generated by the iterative diagonalizations. Here $X=K$ and $X=D$ denote the kept states and the discarded states, respectively. 
In this paper, we focus on the $\sigma_{z}-\sigma_{z}$ correlation function $C(\omega)$, defined as
\begin{eqnarray}
C(\omega)=(1/2\pi)\int_{-\infty}^{\infty} (1/2)\langle[\sigma_{z}(t),\sigma_{z}(0)\rangle]_{+}\rangle e^{i\omega t}dt.
\end{eqnarray} 
We will mainly use the FDM NRG method to calculate $C(\omega)$ at finite temperature. FDM NRG employs the complete basis set composed of discarded states of all NRG shells.\ucite{Anders2,Anders3}
Under this complete basis set, a Lehmann representation for $C(\omega)$ can be written down. By using the NRG approximation
\begin{eqnarray}
H_{N}\vert{se}\rangle_{n}^{X} \approx {E_{s}^{n}}\vert{se}\rangle_{n}^{X},  \,\,\,\,\,\,\,\,\, (X=K, D)
\end{eqnarray} 
the Lehmann representation can be evaluated. In this approximation, the matrix elements of the density operator $\rho = e^{-\beta H_N}/Tr(e^{-\beta H_N})$ is treated exactly. That is, for a given temperature, the contribution from all eigenstates of $\rho$ are taken into account exactly according to the Boltzmann distribution. The excitation energies in the dynamical correlation function, however, are approximated by the intra-shell excitations only. The obtained FDM NRG method\ucite{Weichselbaum1,Weichselbaum2} has the advantage of conserving the sum rule and being computationally efficient, but the positiveness of diagonal spectral function is not guaranteed in certain situations.\ucite{Yang1}
We also compare the results from FE NRG with those from FDM NRG and find no qualitative differences.

In both FDM and FE NRG methods, the delta peaks in the Lehmann representation of $C(\omega)$ are broadened with a log-Gaussian function,\ucite{Bulla3}
\begin{equation}
   \delta(\omega - \omega_n) \rightarrow \frac{e^{-b^2/4}}{b\omega_n\sqrt{\pi}} \exp { \left[ -\frac{( \ln{\omega}-\ln{\omega_n} )^2}{b^2} \right] }.
\end{equation}
Here $b$ is the broadening parameter.

\section{Result}

\subsection{Existence of Thermal Peak at $\omega \sim \omega_T$}
\begin{figure}[t!]   
\begin{center}
\includegraphics[width=480pt, height=360pt,angle=0]{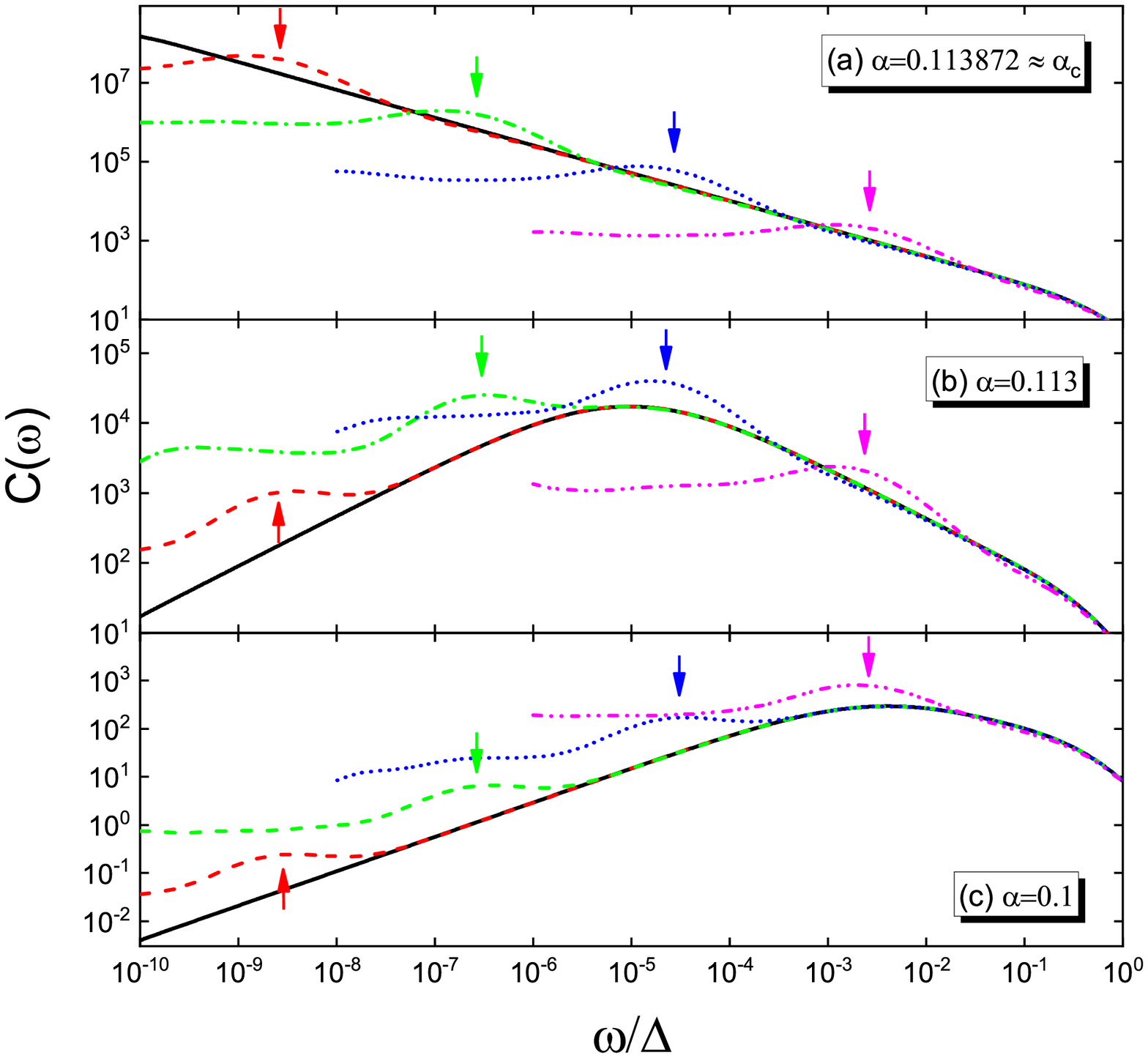}
\end{center}
\parbox[c]{15.0cm}{\footnotesize{\bf Fig.~1.} (color online)
Correlation function $C(\omega)$ at different temperatures. Panels (a)-(c) correspond to $\alpha=0.113872 \approx \alpha_c$, $0.113$, and $0.1$, respectively. In each panel, $T/\Delta = $ $10^{-18} \approx 0$ (black solid line), $10^{-8}$ (red dashed line), $10^{-6}$ (green dash-dotted line), $10^{-4}$ (blue short-dashed line), and $10^{-2}$ (pink dash-dot-dot line). Other parameters are $s=0.7$, $\Delta=0.01$, and $\epsilon=0$. NRG parameters are $N_b=8$, $N=40$, $M_s=100$, $\Lambda=2.0$, and $b=1$. The peak positions are marked by arrows.}
\end{figure}

\begin{figure}[t!]   
\begin{center}
\includegraphics[width=360pt, height=240pt,angle=0]{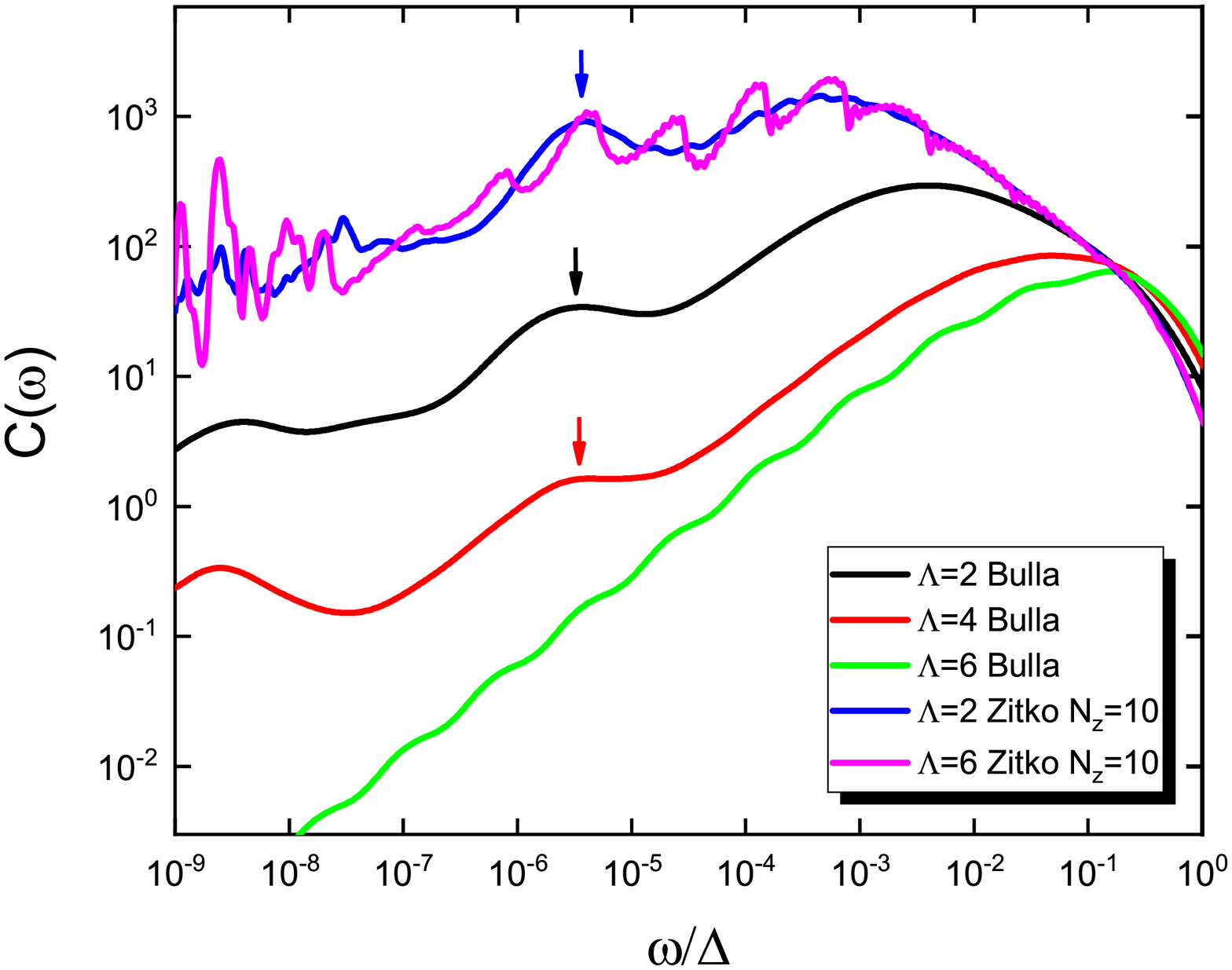}
\end{center}
\parbox[c]{15.0cm}{\footnotesize{\bf Fig.~2.} (color online)
Correlation function $C(\omega)$ at $T/\Delta=10^{-5}$ obtained at different $\Lambda$ values. Different color refers to different $\Lambda$. The lowest three curves are obtained from Bulla's discretization method and using $b=1$. The top two curves are from Zitko's discretization method with $N_z=10$ and $b=1/N_z$. The parameters are $s=0.7$, $\Delta=0.01$, $\epsilon=0$, and $\alpha=0.1$. NRG parameters are $N_b=8$, $N=40$, and $M_s=100$. The arrows mark out the peak positions.}
\end{figure}

For zero temperature, previous studies of $C(\omega)$ show that besides the tunnelling peak at the renormalized tunnelling $\Delta_r$, there is a power-law behaviour $C(\omega)\sim \omega^{s}$ in the low frequency regime $\omega \ll \omega_{cr}$ and critical behavior $C(\omega) \sim\omega^{-s}$ in the intermediate frequency $\omega  \gg \omega_{cr}$.\ucite{Anders1} The crossover frequency $\omega_{cr}$ follows the crossover energy scale $T^{\ast} \sim (\alpha_c - \alpha)^{z\nu}$\ucite{Bulla1} between the delocalized fixed point and the critical fixed point. Here $z$ and $\nu$ are the dynamical and correlation function exponents, respectively. 

For finite temperature, the properties of $C(\omega)$ is less clear. In Fig.1, we focus on the sub-Ohmic bath $s=0.7$ and the nonbiased case $\epsilon=0$. Representative results are produced by FDM NRG using a set of typical NRG parameters $\Lambda=2.0$, $N_b=8$, $M_s=100$ and $N=40$. Panel (a) is for $\alpha = 0.113872 \approx \alpha_c$. Panels (b) and (c) correspond to two values of $\alpha$ in the delocalized phase. 
Each panel contains $C(\omega)$ of four finite temperatures $T/\Delta=10^{-8}$, $10^{-6}$, $10^{-4}$, and $10^{-2}$. These curves are compared with the zero temperature curve obtained at $T/\Delta=10^{-18}$.

For a fixed $\alpha$, it is seen that the high frequency part of $C(\omega)$ does not change much with temperature. As the frequency decreases to a temperature-dependent frequency $\omega_{T} \propto T$ (marked by arrows in Fig.1), $C(\omega)$ deviates from the zero temperature curve, forming a broad peak. For even lower frequency $\omega \ll \omega_{T}$, $C(\omega)$ decreases a bit and a plateau is formed. It does not follow the power forms $C(\omega) \sim \omega^{\pm s}$ any more. This is quite different from the zero temperature case where $C(\omega) \propto \omega^{s}$ down to $T=0$. This scenario holds for different $\alpha$ values, whether it is close to the critical point (Fig.1(a)), in weak coupling (Fig.1(c), or in intermediate coupling (Fig.1(b)).
In Fig.1(b), for different $T$, cases of $\omega_{T} > \omega_{cr}$ as well as $\omega_{T} < \omega_{cr}$ appear. Apparently if $\omega_{T} > \omega_{cr}$, the thermal peak will takeover the low frequency behavior such that the crossover at $\omega_{cr}$ will not occur. But if $\omega_{T} < \omega_{cr}$, the thermal peak will not influence the crossover peak at $\omega_{cr}$.

Below, we check that the above observation of a peak in $C(\omega)$ at $\omega_T$ is not the artefact of NRG errors.
First, we study the influence of the discretization parameter $\Lambda$ on the results. It is known that the discretization error could distort the NRG results for $\Lambda > 1$. In Fig.2, we compare the $C(\omega)$ curve of $T/\Delta =10^{-5}$ obtained at $\Lambda=6.0$, $4.0$, and $2.0$ using Bulla's discretization method, originally developed by Wilson\ucite{Wilson1, Bulla4}. For these calculations, we use broadening parameter $b=1.0$. We also plot two curves obtained at $\Lambda=2.0$ and $\Lambda=6.0$ using Zitko's discretization method (upper two curves in Fig.2). 

Zitko's discretization method\ucite{Zitko1} requires that the hybrid function after the discretization be equivalent to the original hybrid function. It is supposed to lead to much less discretization error than Bulla's method. For the two curves from Zitko's method, the z-average trick\ucite{Yoshida1,Oliveira1,Campo1} is used to further reduce the effect of broadening. Specifically, the result is obtained by averaging $N_z=10$ curves, each obtained with slightly twisted boundaries on the energy axis in the discretization process and with a much smaller broadening parameter $b=1/10$.

In Fig.2, apart from the curve for $\Lambda=6.0$ with Bulla's discretization, all curves have a peak at a common $\omega_T$. For Bulla's discretization, with decreasing $\Lambda$, $C(\omega)$ shifts upwards and the peak at $\omega_{T}$ gets more pronounced. The two curves from Zitko's discretization method stay on the top of the figure and almost coincide with each other, implying that they represent the converged curve at $\Lambda=1.0$. The curve for $\Lambda=6.0$ has too much discretization error such that the thermal peak at $\omega_{T}$ does not appear.

\begin{figure}[t!]   
\begin{center}
\includegraphics[width=360pt, height=280pt,angle=0]{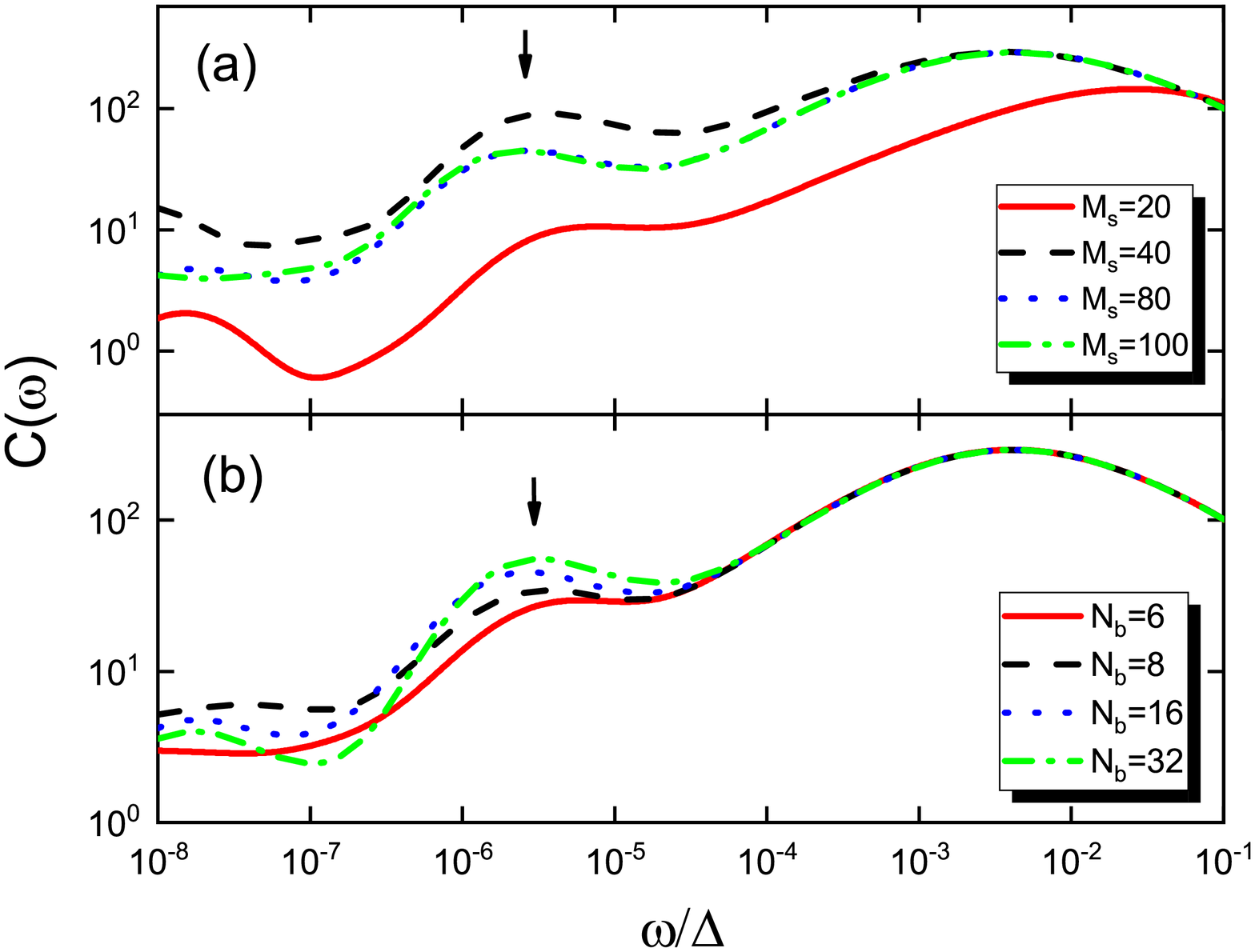}
\end{center}
\parbox[c]{15.0cm}{\footnotesize{\bf Fig.~3.} (color online)
Correlation function $C(\omega)$ for (a) different $M_s$ at $N_b=16$, and (b) different $N_b$ at $M_s = 80$. The parameters are $s=0.7$, $\Delta=0.01$, $\epsilon=0$, $\alpha=0.1$, $T/{\Delta}=10^{-5}$. The other NRG parameters are $N=40$, $\Lambda=2$, and $b=1$. The peak positions are marked out by arrows.}
\end{figure}

Second, we study the influence of NRG truncation error on $C(\omega)$. Recently, the NRG truncation error has been analysed for the static physical quantities.\ucite{Yang2} Systematic analysis of the NRG truncation error in dynamical quantities is still missing. In Fig.3(a), $C(\omega)$ curves from several $M_s$ are compared at $N_b=16$. It is seen that for $\Lambda=2.0$ and $N_b=16$, $M_s=80$ and $M_s=100$ give almost identical $C(\omega)$ in the regime $\omega \gtrsim \omega_{T}$, showing that $Ms =100$ is sufficient to give converged result for this regime. But the low frequency regime $\omega \lesssim \omega_{T}$ is much more difficult to converge with increasing $M_s$. 

In Fig.3(b), we show the influence of $N_b$ on $C(\omega)$ using $M_s=80$. With increasing $N_b$, $C(\omega)$ does not change in  the regime $\omega \gg \omega_T$ but the peak at $\omega = \omega_T$ becomes more pronounced. This supports that the peak at $\omega_T$ exists even in the limit of $N_b=\infty$. Similar as the dependence on $M_s$, $C(\omega)$ in the regime $\omega \lesssim \omega_T$ depends irregularly on $N_b$, making it difficult to draw a reliable conclusion on the behavior of $C(\omega)$ in the low frequency limit. 
Combining Fig.2 and Fig.3, we draw the conclusion that at finite temperature, $C(\omega)$ has a peak at $\omega_T$ which is proportional to temperature. $C(\omega)$ merges with the zero temperature curve above this frequency while deviates significantly from the power-law form $\omega^{\pm s}$ below this frequency.

\subsection{$C(\omega)$ for various $\alpha$ and $s$ values}

In Fig.4, we present $C(\omega)$ at $T/\Delta=10^{-5}$ for various values of $\alpha \leqslant \alpha_c$ and at $\epsilon=0.0$. For very small $\alpha$ values, i.e., for $\alpha \ll \alpha_c$, it can be seen that a tunnelling peak at $\omega = \Delta$ is directly connected to the $\omega^{s}$ power-law regime at lower frequency. For $\alpha \lesssim \alpha_c$, the tunnelling peak gets smeared and a clear crossover peak appears at $\omega_{cr}$, which separates the $\omega^{-s}$ behavior in the higher frequency regime from the $\omega^{s}$ behavior in the lower frequency regime. See the curve for $\alpha=0.1$. At $\alpha = \alpha_c$, the critical behavior $\omega^{-s}$ would extend to the low frequency limit if $T=0$. In all these curves for $T/\Delta=10^{-5}$, the thermal peak at $\omega_T$ are present, with almost $\alpha$-independent $\omega_T$ values. Note that for $\alpha$ very close $\alpha_c$, $\omega_{cr} <  \omega_{T}$ will occur and the thermal peak at $\omega_T$ and the irregular $C(\omega)$ behavior in low frequency regime will takeover. In that case the crossover at $\omega_{cr}$ is no longer present. See the curves for $\alpha=0.113$ and $0.114$.

\begin{figure}[t!]   
\begin{center}
\includegraphics[width=360pt, height=240pt,angle=0]{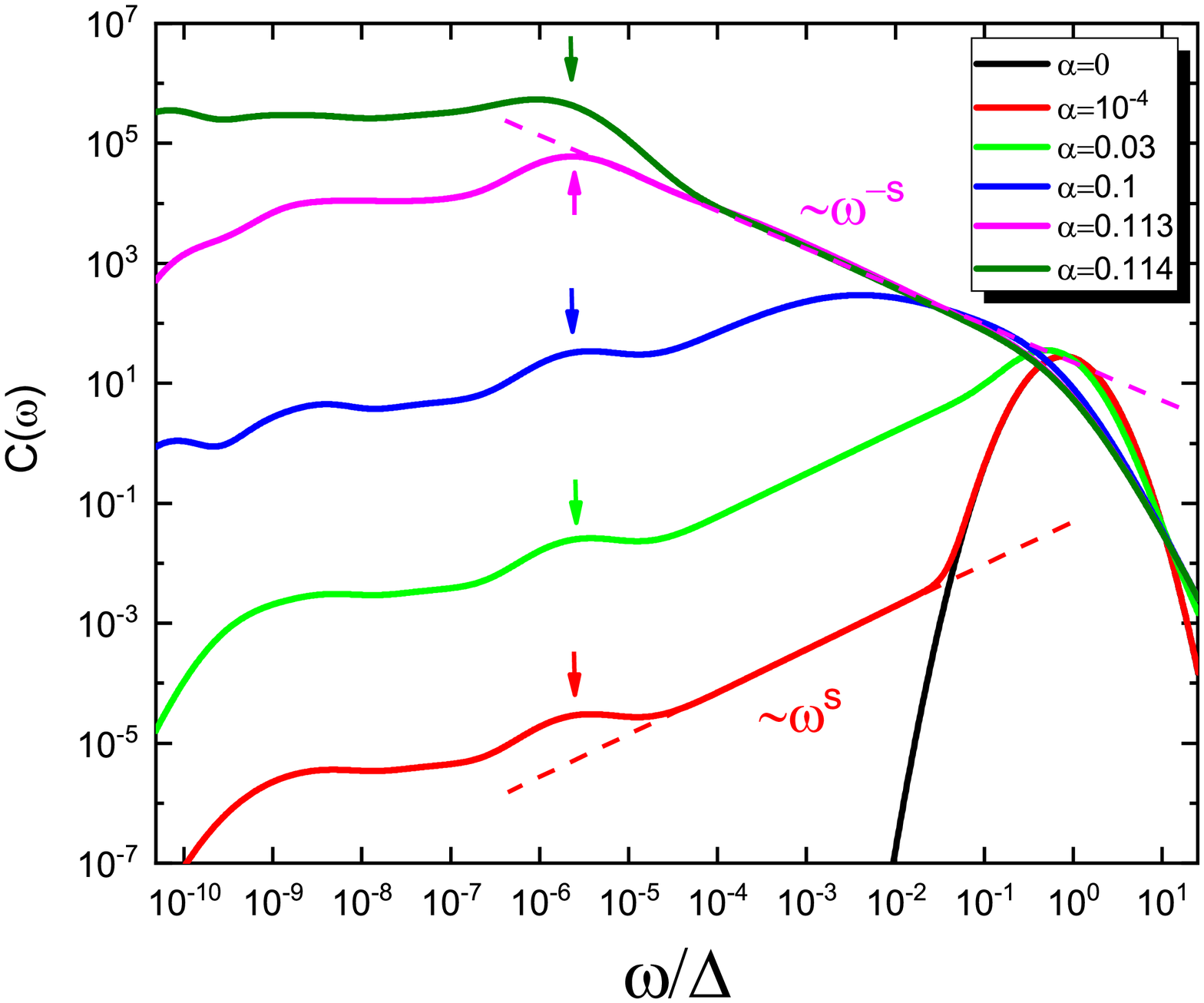}
\end{center}
\parbox[c]{15.0cm}{\footnotesize{\bf Fig.~4.} (color online)
Correlation function $C(\omega)$ at $T/{\Delta}=10^{-5}$ for various $\alpha$ values. From bottom to top in the low frequency regime, $\alpha =0.0$, $10^{-4}$, $0.03$, $0.1$, $0.113$, and $0.114 \approx \alpha_c$, respectively. Other parameters are $s=0.7$, $\Delta=0.01$ and $\epsilon=0$. The NRG parameters are $N_b=8$, $N=40$, $M_s=100$, $\Lambda=2$, and $b=1$. The peak positions are marked out by arrows. The dashed lines show $\omega^{s}$ and $\omega^{-s}$ behavior. }
\end{figure}

\begin{figure}[t!]   
\begin{center}
\includegraphics[width=360pt, height=240pt,angle=0]{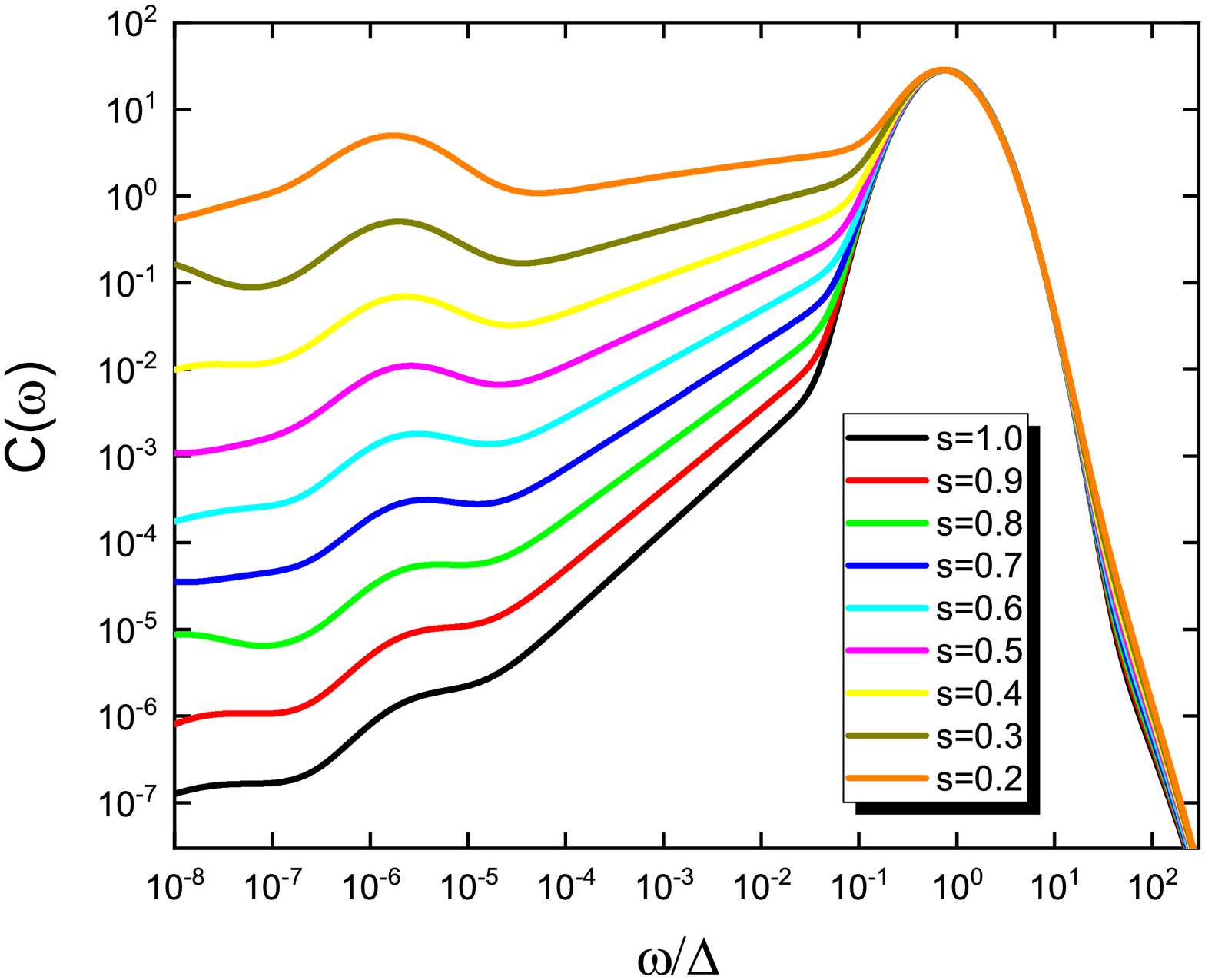}
\end{center}
\parbox[c]{15.0cm}{\footnotesize{\bf Fig.~5.} (color online)
Correlation function $C(\omega)$ at $T/\Delta=10^{-5}$ for various $s$ values. Other parameters are  $\Delta=0.01$, $\epsilon=0$, and $\alpha=0.001$. The NRG parameters are $N_b=8$, $N=40$, $M_s=100$, $\Lambda=2$, and $b=1$. }
\end{figure}

Fig.5 shows the evolution of $C(\omega)$ curve with $s$ increasing from $0.1$ to $1.0$, with fixed $\alpha=0.001$ which is in the delocalized phase for all $s$ values shown in the figure. A robust tunnelling peak at $\omega = \Delta$ is always present through this process. The low frequency regime has $c \omega^{s}$ behavior. Besides the change of slope in the log-log plot in the low frequency regime, the coefficient $c$ decreases with increasing $s$. This is mainly because as $s$ increases, $\alpha_c$ increases. This makes the state at $\alpha = 0.001$ further away from the critical point and the coupling strength is effectively reduced. In all these curves, there are thermal peaks at $\omega_T$ with an $s$-independent $\omega_T$ value. The curves in $\omega < \omega_T$ does not show a persistent pattern with $s$.

\subsection{$C(\omega)$ in $\omega \ll \omega_{T}$ regime}

\begin{figure}[t!]   
\begin{center}
\includegraphics[width=360pt, height=240pt,angle=0]{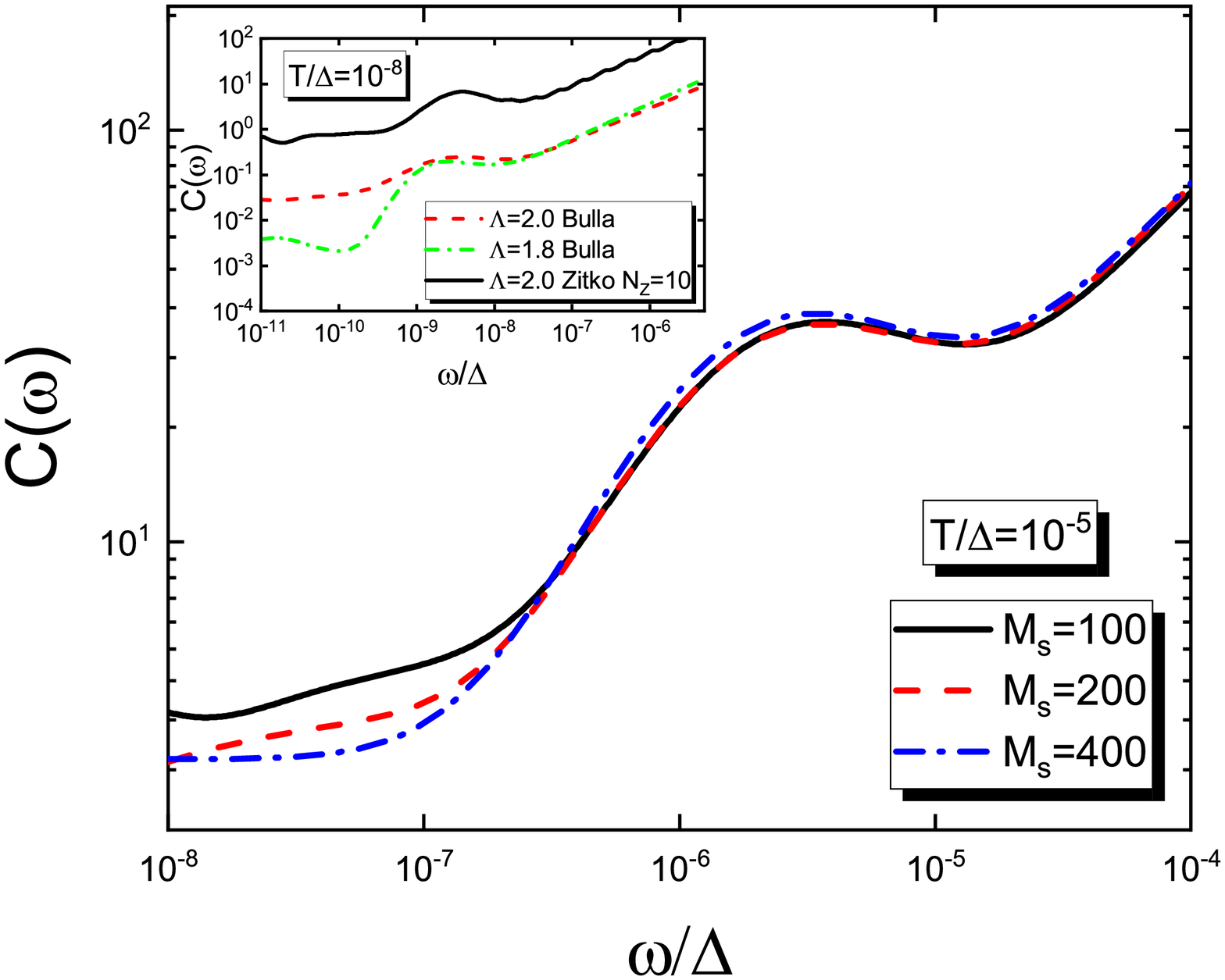}
\end{center}
\parbox[c]{15.0cm}{\footnotesize{\bf Fig.~6.} (color online)
The low frequency part of $C(\omega)$ at $T/{\Delta}=10^{-5}$ for different $M_s$ values. The parameters are $s=0.7$, $\Delta=0.01$, $\epsilon=0$, and $\alpha=0.1$. The NRG parameters are $N_b=8$, $N=40$, $M_s=100$, $\Lambda=2$, and $b=1$. Inset: low frequency part of $C(\omega)$ at $T/\Delta=10^{-8}$ for different $\Lambda$ values. }
\end{figure}

Finally, we discuss our exploration into the low frequency regime of $C(\omega)$, i.e., in $\omega \ll \omega_T$. In Fig.6, we plot $C(\omega)$ curves for different $M_s$ values (main figure) and $\Lambda$ values (inset). Although the thermal peak at $\omega_T$ is robust for all variation of these parameters, $C(\omega)$ in the lower frequency regime $\omega \ll \omega_T$ does not show the signature of convergence. In the main figure of Fig.6, it seems that as $M_s$ increases, the low frequency part of $C(\omega)$ continues to decrease, leading to a pseudo gap behavior below $\omega_T$. This observation, however, is not conclusive given the variation of available data. The curve of $C(\omega)$ always seems to be irregular in $\omega \ll \omega_T$ such that we have not found a persistent pattern in the low frequency limit in the large $M_s$ limit. 
In the inset, the same situation appears for decreasing $\Lambda$ towards unity. Although we have not understood this phenomena completely, we suspect that this loss of descriptive power of NRG for dynamical correlation functions in the regime $\omega \ll T$ is not an accidental phenomena, but a general problem rooted in the NRG algorithm. This problem was observed in the patching method for computing dynamical correlation functions. Although it is claimed that FDM NRG can solve this problem for Anderson impurity model,\ucite{Weichselbaum1} here we find that for SBM the problem still exists for FDM NRG.

To elucidate this issue, we compare the $C(\omega)$ curves from FDM NRG and FE NRG in Fig.7. FE NRG fully takes into account all excitations in $C(\omega)$, including both the intra-shell ones and the inter-shell ones.\ucite{Yang1} It is expected to be more accurate than the FDM NRG which only takes into account the intra-shell excitations in $C(\omega)$. The curves of $C(\omega)$ at two different temperatures show same feature. Both methods produce almost identical curves in the high frequency regime $\omega > \omega_T$. For $\omega \lesssim \omega_T$, the curves from FDM and FE NRG methods deviate from each other, although both have the thermal peak at same $\omega_T$. In the low frequency regime, they differ significantly, with FE result always larger than FDM results. In the large $M_s$ limit, both methods becomes exact and they should produce identical results. This consistence of two methods has not been reached by our present computation. This shows that FE NRG also suffers the same loss of accuracy in $\omega \ll T$. The in depth discussion of this issue is left for a future study.

\begin{figure}[t!]   
\begin{center}
\includegraphics[width=380pt, height=280pt,angle=0]{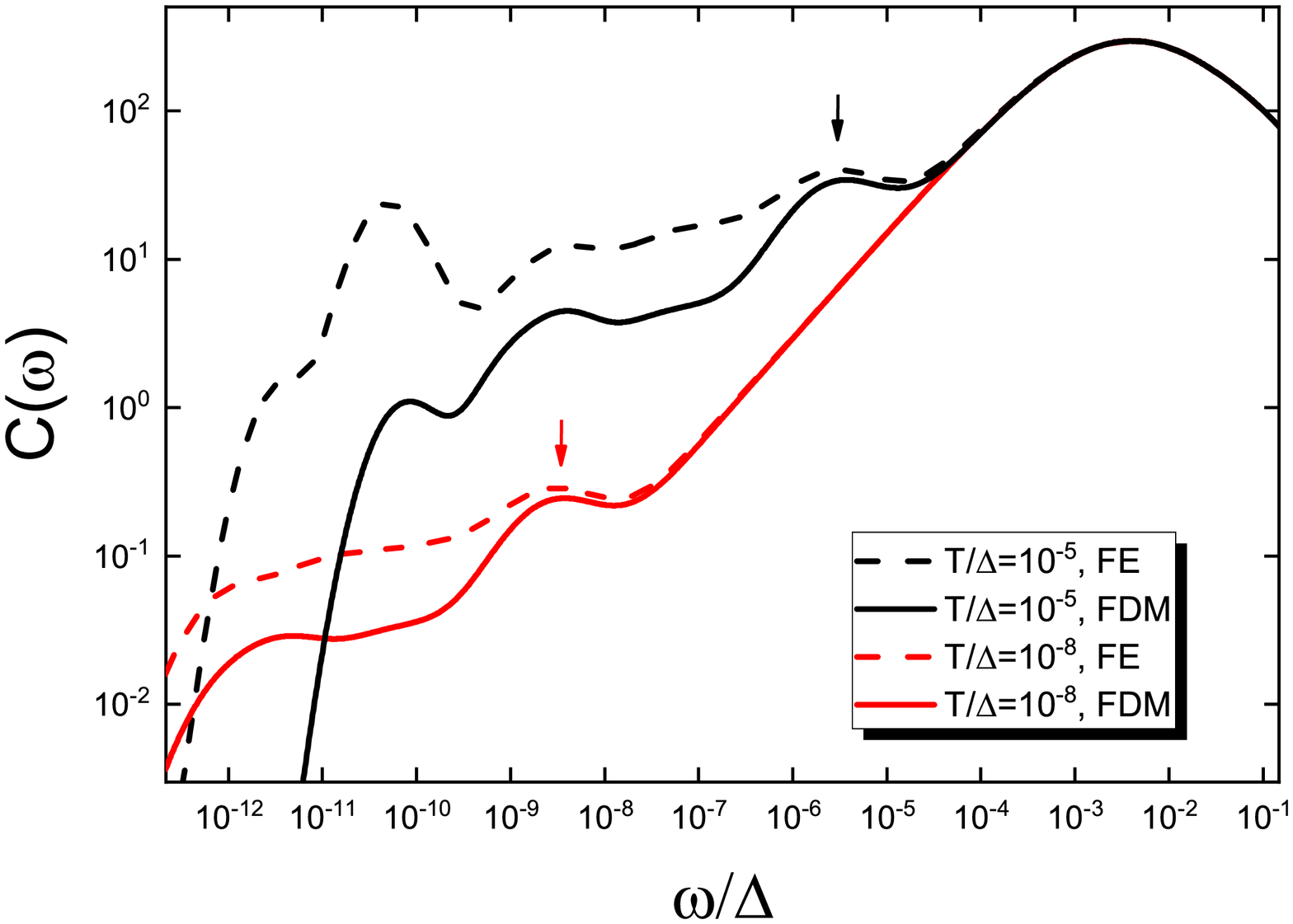}
\end{center}
\parbox[c]{15.0cm}{\footnotesize{\bf Fig.~7.} (color online)
Comparison of $C(\omega)$ from FDM NRG and FE NRG. 
The parameters are $s=0.7$, $\Delta=0.01$, $\epsilon=0$, and $\alpha=0.1$. The NRG parameters are $N_b=8$, $N=40$, $M_s=100$, $\Lambda=2$, and $b=1$. Peak positions are marked our by arrows.}
\end{figure}

\section{Conclusion}

We study the $\sigma_{z}-\sigma_{z}$ correlation function $C(\omega)$ at finite temperature for SBM. A thermal peak is observed at the frequency $\omega_T \sim T$ in the delocalized phase and zero bias. We find that $\omega_T$ is controlled solely by temperature and it is independent of $\alpha$ and $s$. Above this frequency, $C(\omega)$ is almost identical to the zero temperature curve. Below $\omega_T$, $C(\omega)$ significantly deviates from the power-law form $\omega^{\pm s}$ of the zero temperature curve and our NRG calculation gives irregular behavior. A definite conclusion on the behavior of $C(\omega)$ in this regime is yet to be obtained.


\addcontentsline{toc}{chapter}{Acknowledgment}
\section*{Acknowledgment}
NHT acknowledges helpful discussions and share of the Green's function equation of motion results of $C(\omega)$ from Zhiguo L\"u.

\end{CJK*}  


\addcontentsline{toc}{chapter}{References}
\begin{thebibliography}{99}\footnotesize
\itemsep=-3pt plus.2pt minus.2pt   

\bibitem{Leggett1} Caldeira A O and Leggett A J \href{https://doi.org/10.1016/0003-4916(83)90202-6}{1983 \emph{Ann. Mod. Phys.} \textbf{149} 374}

\bibitem{Leggett2} Leggett A J, Chakravarty S, Dorsey A T, Fisher M P A, Garg A, and Zwerger W \href{https://doi.org/10.1103/RevModPhys.59.1}{1987 \emph{Rev. Mod. Phys.} \textbf{59} 1}

\bibitem{Weiss1}
Weiss U 1993 \emph{Quantum Dissipative Systems} (World Scientific, Singapore)

\bibitem{Alex1} Chin A and Turlakov M \href{https://doi.org/10.1103/PhysRevB.73.075311}{2006 \emph{Phys. Rev. B} \textbf{73} 075311}

\bibitem{Anders1} Anders F B, Bulla R, and Vojta M \href{https://doi.org/10.1103/PhysRevLett.98.210402}{2007 \emph{Phys. Rev. Lett.} 
\textbf{98} 21042}

\bibitem{Costi1} Costi T A and McKenzie R H \href{https://doi.org/10.1103/PhysRevA.68.034301}{2003 \emph{Phys. Rev. A} \textbf{68} 034301}

\bibitem{Khveshchenko1} Khveshchenko D V \href{https://doi.org/10.1103/PhysRevB.69.153311}{2004 \emph{Phys. Rev. B} \textbf{69} 153311}

\bibitem{Thorwart1} Thorwart M and Hanggi P \href{https://doi.org/10.1103/PhysRevA.65.012309}{2001 \emph{Phys. Rev. A} \textbf{65} 012309}

\bibitem{Ruokola1} Ruokola B and Ojanen T \href{https://doi.org/10.1103/PhysRevB.83.045417}{2011 \emph{Phys. Rev. B} \textbf{83} 045417}

\bibitem{Chin2} Chin A, Prior J, Huelga S F, and  Plenio M B \href{https://doi.org/10.1103/PhysRevLett.107.160601}{2011 \emph{Phys. Rev. Lett.} \textbf{107} 160601}

\bibitem{Lv1} L\"u Z G and Zheng H \href{https://doi.org/10.1103/PhysRevB.75.054302}{2007 \emph{Phys. Rev. B} \textbf{75} 054302}

\bibitem{Bulla1} Bulla R, Tong N H, and Vojta M \href{https://doi.org/10.1103/PhysRevLett.91.170601}{2003 \emph{Phys. Rev. Lett.} \textbf{91} 170601}

\bibitem{Bulla2} Bulla R, Lee H J, Tong N H, and Vojta M \href{https://doi.org/10.1103/PhysRevB.71.045122}{2005 \emph{Phys. Rev. B} 
\textbf{71} 045122}

\bibitem{Winter1} Winter A, Rieger H, Vojta M, and Bulla R  \href{https://doi.org/10.1103/PhysRevLett.102.030601} {2009 \emph{Phys. Rev. Lett.} \textbf{102} 030601}

\bibitem{Alvermann1} Alvermann A and Fehske H \href{https://doi.org/10.1103/PhysRevLett.102.150601} {2009 \emph{Phys. Rev. Lett.}\textbf{102} 150601}

\bibitem{Kehrein1} Kehrein S K and Mielke A \href{https://doi.org/10.1016/0375-9601(96)00475-6} {1996 \emph{Phys. Lett. A }\textbf{219} 313}

\bibitem{Herbert1} Herbert S and Wilhelm Z \href{https://doi.org/10.1023/A:1004595419419} {1999 \emph{J. Stat. Phys.  }\textbf{94} 1037}

\bibitem{Florens1}Florens L, Freyn A, Venturelli D, and Narayanan R \href{https://doi.org/10.1103/PhysRevB.84.155110}{2011 \emph{Phys. Rev. B} \textbf{84} 155110}

\bibitem{Zheng1} Zheng D C, Wan L, and Tong N H \href{https://doi.org/10.1103/PhysRevB.98.115131}{2016 \emph{Phys. Rev. B} \textbf{98} 115131}

\bibitem{Zheng2} Zheng D C and Tong N H \href{ https://doi.org/10.1088/1674-1056/26/6/060502}{2017 \emph{Chin. Phys. B} \textbf{26} 060502}

\bibitem{Zheng3} Zheng D C and Tong N H \href{ https://doi.org/10.1088/1674-1056/26/6/060501}{2017 \emph{Chin. Phys. B} \textbf{26} 060501}

\bibitem{Sassetti1} Sassetti M and Weiss U \href{https://doi.org/10.1103/PhysRevLett.65.2262}{1990 \emph{Phys. Rev. Lett.} \textbf{65} 2262}

\bibitem{Bulla3} Bulla R, Costi T A, and Vollhardt D \href{https://doi.org/10.1103/PhysRevB.64.045103}{2001 \emph{Phys. Rev. B} \textbf{64} 045103}

\bibitem{Hofstetter1} Hofstetter W \href{https://doi.org/10.1103/PhysRevLett.85.1508}{2000 \emph{Phys. Rev. Lett.} \textbf{85} 1508}

\bibitem{Weichselbaum1} Weichselbaum A \href{https://doi.org/10.1103/PhysRevB.86.245124}{2012 \emph{Phys. Rev. B} \textbf{86} 245124}

\bibitem{Anders2} Anders F B and Schiller S \href{https://doi.org/10.1103/PhysRevLett.95.196801}{2005 \emph{Phys. Rev. Lett.} \textbf{95} 196801}

\bibitem{Anders4} Anders F B and Schiller S \href{https://doi.org/10.1103/PhysRevB.74.245113}{2006 \emph{Phys. Rev. B} \textbf{74} 245113}

\bibitem{Weichselbaum2} Weichselbaum A and Delft J V \href{https://doi.org/10.1103/PhysRevLett.99.076402}{2006 \emph{Phys. Rev. Lett.} \textbf{99} 076402}

\bibitem{Yang1} Yang K and Tong N H \href{https://doi.org/10.1103/PhysRevB.102.085125}{2020 \emph{Phys. Rev. B} \textbf{102} 085125}

\bibitem{Yang2} Yang K and Tong N H arXiv:2010.04398v1 [cond-mat.str-el] 

\bibitem{Wilson1} Wilson K G \href{https://doi.org/10.1103/RevModPhys.47.773}{1975 \emph{Rev. Mod. Phys.} \textbf{47} 773}

\bibitem{Bulla4} Bulla R, Costi T A, and Pruschke T  \href{https://doi.org/10.1103/RevModPhys.80.395 }{2008 \emph{Rev. Mod. Phys.} \textbf{80} 395}

\bibitem{Anders3} Anders F B and Schiller S \href{https://doi.org/10.1103/PhysRevB.74.245113}{2006 \emph{Phys. Rev. B} \textbf{74} 245113}

\bibitem{Zitko1} Zitko R and Pruschke T \href{https://doi.org/10.1103/PhysRevB.79.085106}{2009 \emph{Phys. Rev. B} \textbf{79} 085106}

\bibitem{Yoshida1} Yoshida M, Whitaker M A, and Oliveira L N \href{https://doi.org/10.1103/PhysRevB.41.9403}{1990 \emph{Phys. Rev. B} \textbf{41} 9403}

\bibitem{Oliveira1} Oliveira W C and Oliveira L N \href{https://doi.org/10.1103/PhysRevB.49.11986}{1990 \emph{Phys. Rev. B} \textbf{49} 11986}

\bibitem{Campo1}  Campo V L and Oliveira L N\href{https://doi.org/10.1103/PhysRevB.72.104432}{2005 \emph{Phys. Rev. B} \textbf{72} 104432}



\end{thebibliography}
\end{document}